\documentclass[conference]{IEEEtran}
\IEEEoverridecommandlockouts
\usepackage{cite}
\usepackage{multirow}
\usepackage{booktabs}
\usepackage{amsmath,amssymb,amsfonts}
\usepackage{algorithmic}
\usepackage{graphicx}
\usepackage{textcomp}
\usepackage{xcolor}
\usepackage{hyperref}	

\newcommand\blfootnote[1]{%
  \begingroup
  \renewcommand\thefootnote{}\footnote{#1}%
  \addtocounter{footnote}{-1}%
  \endgroup
}

\usepackage{algorithm}
\usepackage{amsmath}
\usepackage{amsfonts}

\usepackage{authblk}

\hypersetup{colorlinks,
	linkcolor=blue,%
        anchorcolor=blue,
        urlcolor=blue,
	citecolor=blue}

\def\BibTeX{{\rm B\kern-.05em{\sc i\kern-.025em b}\kern-.08em
    T\kern-.1667em\lower.7ex\hbox{E}\kern-.125emX}}
\begin{document}

\title{An Intra- and Cross-frame Topological Consistency Scheme for Semi-supervised Atherosclerotic Coronary Plaque Segmentation}

\author[1, $\dagger$]{Ziheng Zhang}
\author[3, $\dagger$]{Zihan Li}
\author[4]{Dandan Shan}
\author[1]{Yuehui Qiu} 
\author[1,2,4 *]{Qingqi Hong}
\author[1,4 *]{Qingqiang Wu}
\affil[1]{Center for Digital Media Computing, School of Film, School of Informatics, Xiamen University, Xiamen, China}
\affil[2]{National Institute for Data Science in Health and Medicine, Xiamen University, Xiamen, China}
\affil[3]{University of Washington, Seattle, USA}
\affil[4]{Institute of Artificial Intelligence, Xiamen University, Xiamen, China}


\maketitle

\begin{abstract}
Enhancing the precision of segmenting coronary atherosclerotic plaques from CT Angiography (CTA) images is pivotal for advanced Coronary Atherosclerosis Analysis (CAA), which distinctively relies on the analysis of vessel cross-section images reconstructed via Curved Planar Reformation. This task presents significant challenges due to the indistinct boundaries and structures of plaques and blood vessels, leading to the inadequate performance of current deep learning models, compounded by the inherent difficulty in annotating such complex data. To address these issues, we propose a novel dual-consistency semi-supervised framework that integrates Intra-frame Topological Consistency (ITC) and Cross-frame Topological Consistency (CTC) to leverage labeled and unlabeled data. ITC employs a dual-task network for simultaneous segmentation mask and Skeleton-aware Distance Transform (SDT) prediction, achieving similar prediction of topology structure through consistency constraint without additional annotations. Meanwhile, CTC utilizes an unsupervised estimator for analyzing pixel flow between skeletons and boundaries of adjacent frames, ensuring spatial continuity. Experiments on two CTA datasets show that our method surpasses existing semi-supervised methods and approaches the performance of supervised methods on CAA. In addition, our method also performs better than other methods on the ACDC dataset, demonstrating its generalization. \blfootnote{$\dagger$ means equal contribution; * Corresponding authors: Qingqi Hong, Qingqiang Wu.}
\end{abstract}


\section{Introduction}
Coronary Atherosclerosis Analysis (CAA) is crucial in diagnosing Coronary Artery Disease (CAD), 
involving tasks such as stenosis classification and plaque identification. However, traditional sparse prediction tasks often fail to provide accurate geometric shapes of plaques. Therefore, we employ the segmentation method to identify plaques accurately, providing physicians with essential information on plaque size and location to aid in assessing CAD-related risks.
Unlike conventional CT image analysis, CAA typically involves Curved Planar Reformation (CPR) \cite{zreik2018recurrent, ma2021transformer, cpr, lin2022deep, qiu2023corsegrec, LiuJiang} to reconstruct a 2D cross-sectional image of the coronary artery, for brevity, let's call it frame image. 
Previous works \cite{zreik2018recurrent, ma2021transformer, yang2021automatic, lin2022deep, HAN, RAJ, SING, Gui2023},  employ fully supervised methods for accomplishing CAA tasks, with no reports about semi-supervised learning in CAA. Labeling CTA or CPR frame images is challenging, which prompts us to adopt a semi-supervised scheme for CAA tasks to leverage unlabeled data. 
Consistency regularization is commonly applied in semi-supervised medical image segmentation. These methods can be roughly grouped into two types: perturbation consistency and task mapping.
The former \cite{MC, CPS, li2023scribblevc, CTCT, li2024scribformer} 
is related to different perturbation methods, and the model must achieve consistent predictions for the perturbed samples or features.
The latter \cite{optic, wang2023swinmm} generates multi-task predictions for different tasks using the original data. These predictions are then placed into a common predefined space, and consistency constraints are formed by reducing the distance between these tasks. This predefined space can be adjusted to fit the requirements of a specific medical study, such as Optic Disc and Cup Segmentation \cite{optic}.
\begin{figure}
\setlength{\abovecaptionskip}{-4.5mm}
\centering
\includegraphics[width=0.48\textwidth]{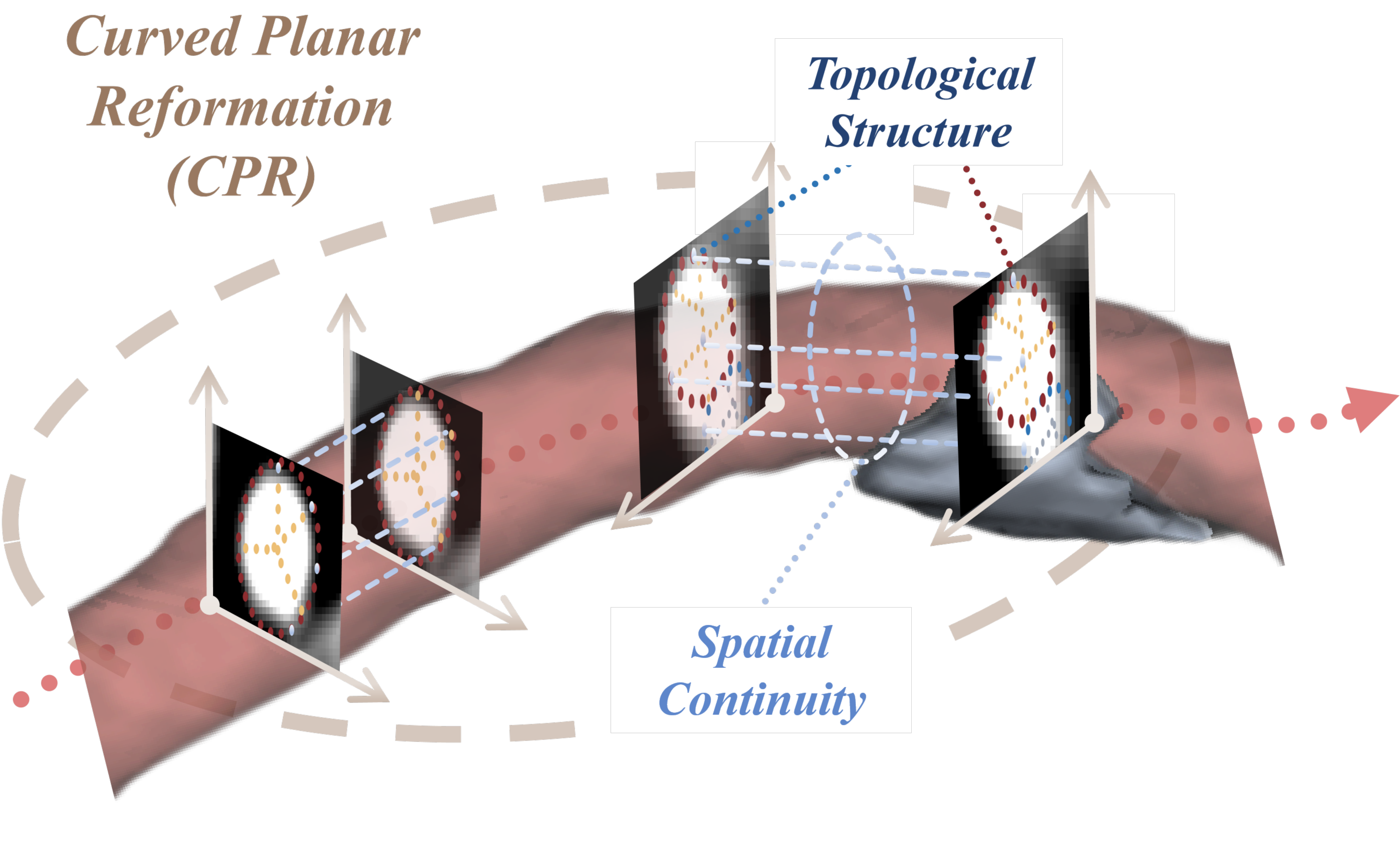}
\caption{CPR Schematic. There is spatial continuity in the frame images, and their topological structure can be used as prior information.} \label{CPR}
\vspace{-6.5mm}
\end{figure}
According to the characteristics of CPR, there is a spatial continuity in the frame images, and its topological structure can be used as prior information, as shown in Fig.\ref{CPR}. Therefore, We propose a dual-consistency semi-supervised framework, incorporating Intra-frame Topological Consistency (ITC) and Cross-frame Topological Consistency (CTC). The main contributions of this paper can be summarized as follows:

\begin{itemize}
\item[$\bullet$] In contrast to fully supervised CAA methods, we propose a novel dual-consistency semi-supervised framework that integrates Intra-frame Topological Consistency (ITC) and Cross-frame Topological Consistency (CTC) to leverage both labeled and unlabeled data. 
\item[$\bullet$] ITC employs a dual-task network for achieving consistent prediction of topology structure through consistency without additional annotations; CTC utilizes an unsupervised estimator for analyzing pixel flow between skeletons and boundaries of adjacent frames, ensuring spatial continuity.
\item[$\bullet$] Experimental results on two CTA datasets show that our method surpasses existing semi-supervised methods and fully-supervised methods on CAA. On the ACDC dataset, our method also performs better than other methods, proving the generalization.
\end{itemize}

\begin{table}[]
\setlength{\abovecaptionskip}{0mm}
\centering
\setlength{\tabcolsep}{0.5mm}
\caption{Summary of all defined variables in the paper}
\label{Variable Summary}
\begin{tabular}{cc}
\hline
\textbf{Variable} & \textbf{Definition} \\ \hline 
$I$ & Input data, contains two adjacent frame images $I_{t,t+1}$ \\  
$Y^{S}$ & Segmentation GT \\ 
$Y^{R}$ & Regression SDT label \\ 
$\hat{S}$ & Segmentation mask prediction \\ 
$\hat{R}$ & Regression SDT prediction\\ 
$\widetilde{S}$ & Pseudo label \\ 
$Z^{K,B}$ & Point set of skeleton/boundary, extracted from the $Y^{R}$ \\
$\hat{Z}^{K,B}$ & Point set of skeleton/boundary, extracted from the $\hat{R}$ \\
$\hat{R}_{t, t+1}$ & SDT prediction of a single frame $I_{t,t+1}$ \\ 
$\sum_{i=1}^{D}\{\hat{R}_{t, t+1}^{i}\}$ & Pyramid features of $\hat{R}_{t, t+1}$ \\ 
$\sum_{i=1}^{D}\{\hat{O}_{f}^{i}\}$ & Multi-scale pixel flow from $\hat{R}_{t}$ to $\hat{R}_{t+1}$ \\ \hline
\end{tabular}
\vspace{-4mm}
\end{table}

\begin{figure*}[!ht] 
\centering
\includegraphics[width=0.9\textwidth]{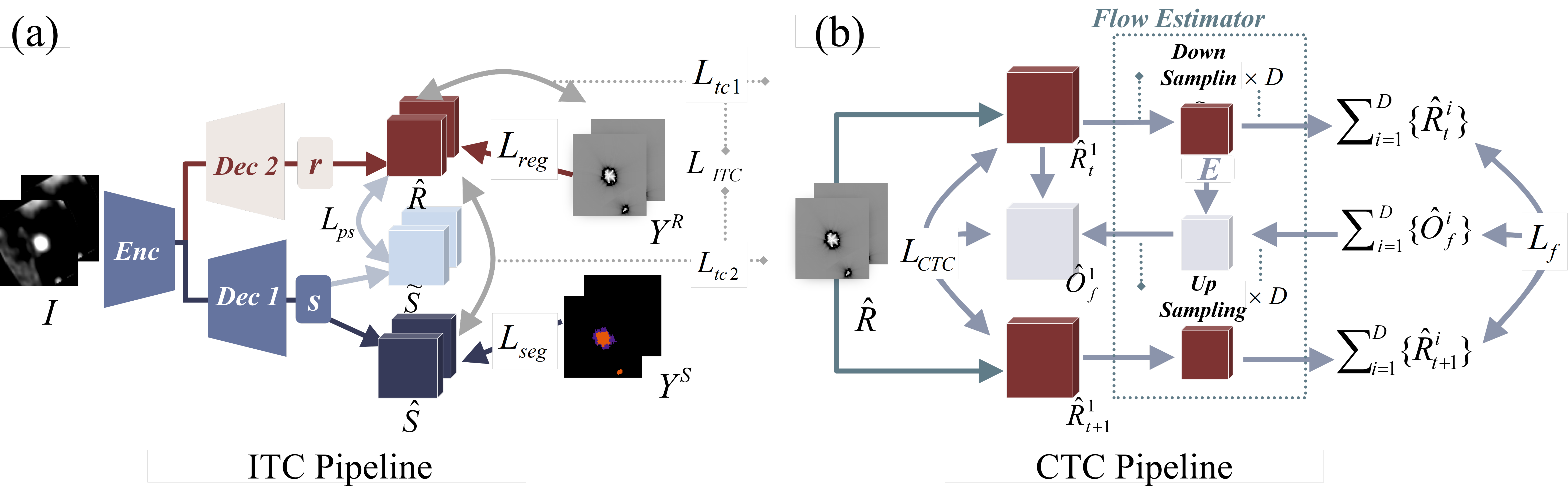}
\vspace{-2mm}
\caption{The overview of our method: \textbf{(a) ITC Pipeline.} Two adjacent frame images share a dual-task network, involving regression and segmentation tasks. We simultaneously construct ITC between the regression results and SDT labels, between the regression results and segmentation results. \textbf{(b) CTC Pipeline.} Input and split the regression results into a single one. Via an unsupervised flow estimator. CTC is constructed on the adjacent regression results with the learned pixel flow.} 
\label{Pipeline}
\vspace{-3mm}
\end{figure*}

\section{Methodology}
Data is divided into labeled data set $D_{l}=\{I\}^{N_l}$ and unlabeled data set $D_{u}=\{I\}^{N_u}$. Each data $I$ contains two adjacent frame images, $I_{t}$ and $I_{t+1}$. 
The labels include segmentation GT: $Y^{S}$;
SDT label: $Y^{R}$ 
, which is from GT, and ignore the background. Other variables that frequently appear in this article can be seen in Table. \ref{Variable Summary}.
\subsection{Intra-frame Topological Consistency}
We use a dual-task network to simultaneously predict the segmentation mask and SDT, as shown in Fig. \ref{Pipeline}(a). This network consists of a U-Net \cite{unet} encoder and two decoders: a U-Net decoder for segmentation mask generation and an MSRF \cite{msrf} decoder for SDT prediction. The U-Net encoder extracts features from the input image, which are processed by the decoders to perform the segmentation and regression tasks. For input $I$, these adjacent frame images share the network. The regression task produces $\hat{R}$, while the segmentation task produces $\hat{S}$. For labeled inputs $D_{l}$, we use Dice loss $L_{seg}$ and L1 loss $L_{reg}$ to train the segmentation and regression tasks. To better utilize unlabeled data, we apply a soft pseudo-label loss $L_{ps}$ \cite{MC} to guide the training of ICTC.
\subsubsection{\textbf{Skeleton/Boundary Topological Constraints}}
if $I \in D_{l}$, we extract the skeleton/boundary positions from $Y^{R}$, denoted as point sets $Z^{K}$ and $Z^{B}$. Point $x$ in $Z^{K}$ and $Z^{B}$, is associated with indicator functions $\mathbb{I}_{K}$ and $\mathbb{I}_{B}$ respectively. The value \cite{SDT} of the boundary on $Y^{R}$ is $0$, and the skeleton is $1$.
\vspace{-0.1mm}
\begin{equation}
\begin{aligned}
\mathbb{I}_{K}(x) &= 
\begin{cases} 
    1, & \text{If } x \text{ labeled as skeleton in } Y^{R}; \\
    0, & \text{Otherwise. }
\end{cases} \\
\mathbb{I}_{B}(x) &=  
\begin{cases} 
    1, & \text{If } x \text{ labeled as boundary in } Y^{R}; \\
    0, & \text{Otherwise. }
\end{cases}
\end{aligned}
\end{equation}
\vspace{-1mm}

if $I \in (D_{l} \cup D_{u})$, we also extract the skeleton/boundary positions from the SDT prediction, denoted as point set $\hat{Z}^{K}$, $\hat{Z}^{B}$. The indicator functions $\mathbb{I}_{K_t}$ and $\mathbb{I}_{B_t}$ also apply to $\hat{Z}^{K}$, $\hat{Z}^{B}$, when we replace $Y^{R}$ by $\hat{R}$. The value of the boundary on $\hat{R}$ is $0$, and the value of the skeleton is higher than $0.8$.

The topological position point set not only serves to calibrate the skeleton/boundary but also mirrors the connectivity \cite{SDT} and topological structure of the object.
For $\hat{S}$ and $\hat{R}$ generated by the dual-task network. We expect the binary mask prediction to exhibit 0 at the edge position and 1 at the skeleton position. Similarly, we anticipate the SDT prediction to be approximately 0 at the edge and 1 at the skeleton positions. Therefore, we introduce the Topological Consistency (ITC) constraints as follows:
\vspace{-1mm}
\begin{equation}
L_{ITC} = L_{tc1} + L_{tc2} \label{ITC}
\end{equation}
Then, $L_{tc1}$ measures the accuracy of the SDT prediction by calculating the discrepancy between the predicted values and the ideal values on the skeleton point set $Z^{K}$ as well as the average prediction value on the boundary point set $Z^{B}$.
\begin{equation}
L_{tc1} = \frac{\sum_{x\in Z^{K}}\left|{\hat{R}(x)-1}\right|}{\left|Z^{K}\right|}  + \frac{\sum_{x\in Z^{B}}\hat{R}(x)}{\left|Z^{B}\right|} \label{TC_1}
\end{equation}
Topological constraints not only applies to $Y^{R}_{t}$ and $\hat{R}_{t}$, but also $\hat{R}_{t}$ and $\hat{S}_{t}$. 
$L_{tc2}$ measures the accuracy of the segmentation mask by calculating the error on the predicted skeleton point set $\hat{Z}^{K}$, and the average prediction value on the boundary point set $\hat{Z}^{B}$.
\begin{equation}
L_{tc2} = \frac{\sum_{x\in \hat{Z}^{K}}\left|{\hat{S}(x)-1}\right|}{\left|\hat{Z}^{K}\right|} + \frac{\sum_{x\in \hat{Z}^{B}}\hat{S}(x)}{\left|\hat{Z}^{B}\right|}  \label{TC_2}
\end{equation}

\begin{table*}[!ht]
\begin{minipage}[c]{0.48\textwidth}
    \centering
    \setlength{\tabcolsep}{1mm}
    \caption{Performance Comparison on the CAA-Seg dataset.} \label{Comparison Tab 1}
    \resizebox{!}{2.1cm}{
        \begin{tabular}{ccccccc}
        \hline
        \multirow{2}{*}{\textbf{Method}} 
        & \multirow{2}{*}{Labeled} & \multirow{2}{*}{Un} & \multicolumn{4}{c}{\textbf{Metric}} \\
        \cmidrule(l){4-7}
        &  & & Dice $\uparrow$ & Jac  $\uparrow$ & HD95 $\downarrow$ & ASD $\downarrow$\\ \hline
        %
        nnUNet \cite{nnu} & 48(100\%) & 0 & 73.69 & 66.64 & 6.12 & 2.43 \\ \hline
        URPC \cite{URPC} & 5(10\%) & 43 & 54.72 & 42.60 & 18.08 & 9.59\\
        SSNet \cite{SS} & 5(10\%) & 43 & 56.53 & 48.85 & 16.74 & 8.06\\
        UAMT \cite{UAMT} & 5(10\%) & 43 & 57.24 & 47.97 & 16.04 & 6.81 \\
        MT \cite{MT} & 5(10\%) & 43 & 59.08 & 50.57 & 13.82 & 6.39 \\
        MCNet+ \cite{MC} & 5(10\%) & 43 & 60.53 & 52.69 & 13.02 & 5.75\\
        CTCT \cite{CTCT} & 5(10\%) & 43 & 60.81 & 53.02 & 9.52 & 4.44\\
        CPS \cite{CPS} & 5(10\%) & 43 & 63.71 & 55.88 & 11.52 & 4.99 \\
        %
        ICTC(Ours) & 5(10\%) & 43 & 74.58 & 68.34 & 7.38 & 2.96\\ \hline
    \end{tabular} 
    }
\end{minipage}\hfill
\begin{minipage}[c]{0.48\textwidth}
    \centering
    \setlength{\tabcolsep}{1mm}
    \caption{Performance Comparison on the AD-CTA dataset.} \label{Comparison Tab 2}
    \resizebox{!}{2.1cm}{
        \begin{tabular}{ccccccc}
        \hline
        \multirow{2}{*}{\textbf{Method}} & \multirow{2}{*}{Labeled}
        & \multirow{2}{*}{Un} & \multicolumn{4}{c}{\textbf{Metric}} \\
        \cmidrule(l){4-7}
        &   & 
        & Dice $\uparrow$ & Jac  $\uparrow$ & HD95 $\downarrow$ & ASD $\downarrow$\\ \hline
        nnUNet \cite{nnu} & 73(100\%) & 0 &  73.15 & 59.20 & 5.35 & 2.05 \\ \hline
        SSNet \cite{SS} & 7(10\%) & 66 & 64.38 & 50.67 & 13.20 & 7.24 \\
        MT \cite{MT} & 7(10\%) & 66 & 65.81 & 52.42 & 10.18 & 4.72 \\
        UAMT \cite{UAMT} & 7(10\%) & 66 & 66.39 & 53.14 & 11.07 & 5.29\\
        MCNet+ \cite{MC} & 7(10\%) & 66 & 66.90 & 52.76 & 11.06 & 4.85 \\
        CTCT \cite{CTCT} & 7(10\%) & 66 & 67.63 & 54.11 & 11.60 & 6.76 \\
        URPC \cite{URPC} & 7(10\%) & 66 & 68.87 & 55.54 & 8.88 & 3.70 \\
        CPS \cite{CPS} & 7(10\%) & 66 & 70.28 & 56.25 & 8.63 & 3.15 \\
        ICTC(Ours) & 7(10\%) & 66 & 73.34 & 59.40 & 5.98 & 2.25 \\ \hline
        \end{tabular} 
    }
\end{minipage}
\vspace{-4mm}
\end{table*}

\subsection{Cross-frame Topological Consistency}
\subsubsection{\textbf{Flow Estimator}}
Inspired by the traditional optical flow learning \cite{uflow, upflow}, we utilize an unsupervised estimator to analyze pixel flow between skeletons and boundaries of adjacent frames following the dual-task network, which can ensure spatial continuity, as shown in Fig. \ref{Pipeline}(b). We split $\hat{R}$ into $\hat{R}_{t}$ and $\hat{R}_{t+1}$, and it aims to learn the pixel flow $\hat{O}_{f}$ between $\hat{R}_{t}$ and $\hat{R}_{t+1}$ via the unsupervised estimator. With $\hat{R}_{t+1}$ as the reference frame and $\hat{R}_{t}$ as the source frame, we introduce a constraint $L_{f}$ to minimize the distance between the reference and warped frames.
\begin{equation}
L_{f} = \sum_{i=1}^{D}  [\hat{R}_{t+1}^{i}-\hat{O}_{f}^{i}(\hat{R}_{t}^{i})]^2
\end{equation}
where, $\sum_{i=1}^{D}\{\hat{R}_{t, t+1}^{i}\}$ denotes the pyramid features of $\hat{R}_{t, t+1}$, and $\hat{R}_{t, t+1}^{1} = \hat{R}_{t, t+1}$. $\sum_{i=1}^{D}\{\hat{O}_{f}^{i}\}$ denotes the multi-scale pixel flow encoded from $\hat{R}_{t}^{i}$. $D$ denotes the layers' number of the pyramid, with $D=3$. Here, through $L_{f}$, the learned pixel flow can ensure the spatial continuity in $\hat{R}_{t,t+1}$ and fine-tune $\hat{R}_{t,t+1}$ to satisfy this continuity as much as possible. 

\subsubsection{\textbf{Topological Continuity Constraints}}
Flow estimator learns the spatial continuity $\hat{O}_{f}$ between two frame images. $L_{f}$ can be considered a continuity constraint on global pixels. Additionally, constraints can be imposed on the skeleton/boundary positions. Our CTC loss is defined as follows:
\begin{equation}
L_{CTC} = L_{f} + \sum_{x\in (\hat{Z}^{K}\cup\hat{Z}^{B})}[\hat{R}_{t+1}(x)-\hat{O}_{f}(\hat{R}_{t}(x))]^2 \label{CTC}
\end{equation}
Where, $\hat{O}_{f}$ denotes the output of the flow estimator, and $\hat{O}_{f} = \hat{O}_{f}^{1}$. CTC constructs the spatial continuity of topological structure in frames. The total loss of our method is as follows:
\begin{equation}
L_{Total} = L_{seg} + L_{reg} + L_{ps} + L_{ITC} + L_{CTC}
\end{equation}

\vspace{-2mm}
\begin{table}[ht]
\caption{Performance Comparison on ACDC dataset.} \label{Comparison Tab ACDC} 
\centering
\setlength{\tabcolsep}{1mm}
\resizebox{!}{2.2cm}{
    \begin{tabular}{ccccccc}
\hline
\multirow{2}{*}{\textbf{Method}} & \multirow{2}{*}{Labeled} & \multirow{2}{*}{Un} & \multicolumn{4}{c}{\textbf{Metric}} \\
\cmidrule(l){4-7}
&   &  & Dice $\uparrow$ & Jac  $\uparrow$ & HD95 $\downarrow$ & ASD $\downarrow$\\ \hline
UAMT \cite{UAMT} & 7(10\%) & 63 & 84.07 & 71.70 & 13.10 & 4.08\\
MT  \cite{MT} & 7(10\%) & 63 & 84.86 & 72.63 & 13.63 & 3.65 \\
URPC \cite{URPC} & 7(10\%) & 63 & 84.96 & 73.35 & 12.63 & 3.40 \\
CTCT \cite{CTCT} & 7(10\%) & 63 & 85.32 & 74.04 & 8.96 & 2.71 \\ 
CPS \cite{CPS} & 7(10\%) & 63 & 86.09 & 76.81 &  7.27 & 2.01 \\
SSNet \cite{SS} & 7(10\%) & 63 & 86.78 & 77.67 & 6.07 & 1.40 \\
MCNet+ \cite{MC} & 7(10\%) & 63 & 87.10 & 78.06 & 6.68 & 2.00 \\
ABD \cite{chi2024adaptive} & 7(10\%) & 63 & 86.34 & 76.79 & 4.72 & 1.62 \\
ICTC(Ours) & 7(10\%) & 63 & 88.22 & 78.72 & 7.48 & 1.58 \\ \hline
\end{tabular}
}
\vspace{-3mm}
\end{table} 

\section{Experiments}
\subsubsection{\textbf{Implementation Details}}
We evaluate our method (ICTC) on three datasets, including (1) CTA datasets: CAA-Seg and AD-CTA; (2) Cardiac dataset: ACDC\footnote{\href{https://www.creatis.insa-lyon.fr/Challenge/acdc/databases.html}{https://www.creatis.insa-lyon.fr/Challenge/acdc/databases.html}}. 
CAA-Seg includes 20 patients, with 60 vascular segments. AD-CTA includes 98 vascular segments from 40 patients. Professional medical workers label the lumen and calcified plaques. 
For the ACDC dataset \cite{bernard2018deep}, we follow the same dataset setup according to the previous works \cite{MC}. 
We compare our method with seven semi-supervised segmentation methods and the fully supervised nnUNet \cite{nnu}. For evaluating the models, we utilize four metrics: Dice, Jaccard, Hausdorff Distance (HD95), and Average Surface Distance (ASD). 

\begin{figure*}[!ht]
\centering
\includegraphics[width=0.8\textwidth]{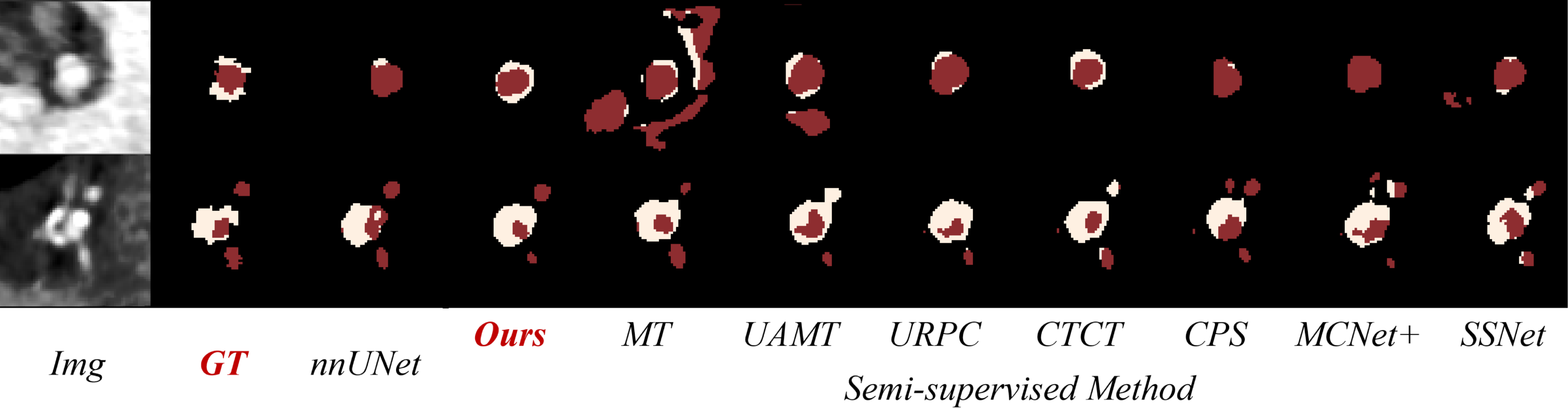}
\vspace{-3mm}
\caption{Visualization of the results on the CAA-Seg. In the figure, the red is the lumen and the white is the calcified plaque.} 
\label{Comparison Vis}
\vspace{-3mm}
\end{figure*}
\begin{figure*}[!ht]
\centering
\includegraphics[width=0.8\textwidth]{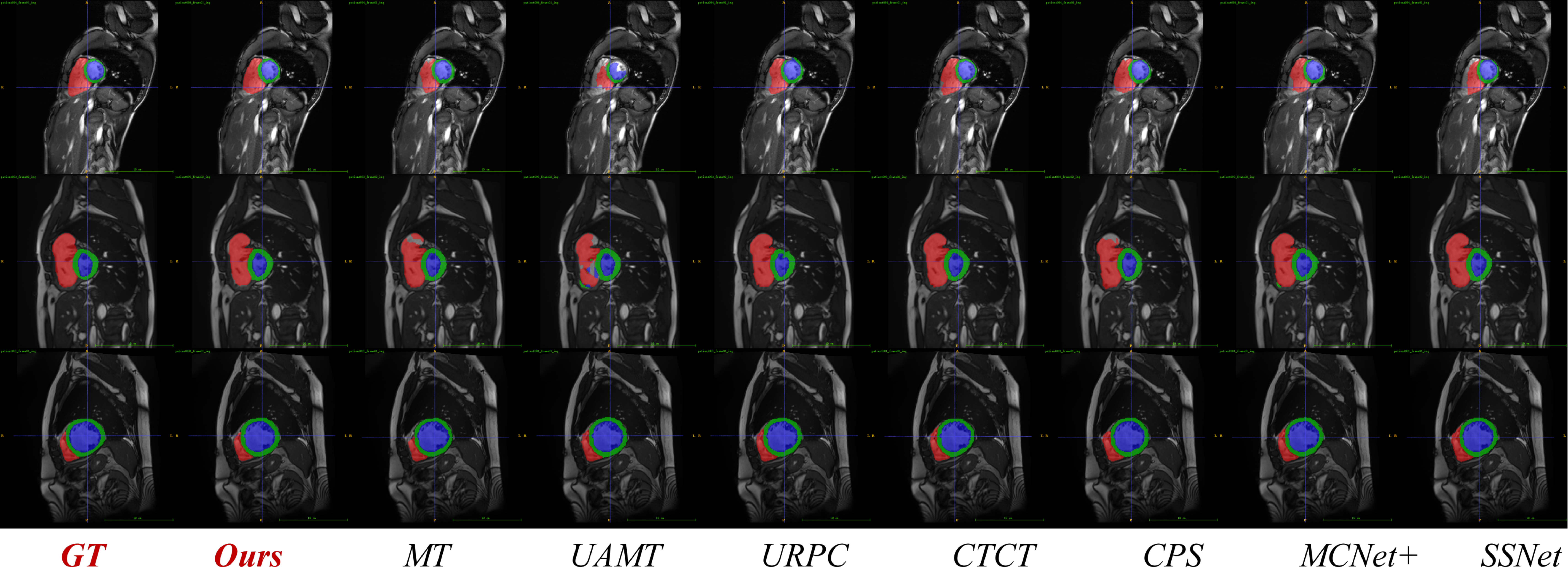}
\vspace{-2mm}
\caption{Visualization of the results on the ACDC. In the figure, the red is the right ventricle (RV), the blue is the left ventricle (LV), and the green is the myocardium.} 
\label{Comparison Vis ACDC}
\vspace{-4mm}
\end{figure*}

\begin{table}[h]
\vspace{-4mm}
\centering
\setlength{\tabcolsep}{0.4mm}
\caption{Ablation studies for the main components on CAA-Seg dataset.} \label{Ablation Tab}
\resizebox{!}{2.1cm}{
    \begin{tabular}{cccccccccc}
    \hline
    \multirow{2}{*}{$Seg$} & \multirow{2}{*}{$Reg$} & \multirow{2}{*}{$L_{ITC}$} & \multirow{2}{*}{$L_{CTC}$}  & \multirow{2}{*}{Labeled} & \multirow{2}{*}{Un} & \multicolumn{4}{c}{\textbf{Metric}} \\
     \cmidrule(l){7-10}     
    & & & & &  & Dice $\uparrow$ & Jac  $\uparrow$ & HD95 $\downarrow$ & ASD $\downarrow$\\ \hline
    \checkmark & & & & 5(10\%) & 0 & 59.16 & 49.67 & 14.56 & 5.83 \\
    \checkmark & \checkmark & &  & 5(10\%) & 0 & 60.08 & 49.85 & 17.51 & 6.68 \\
    \checkmark & \checkmark &  \checkmark &  & 5(10\%) & 0 & 60.31 & 51.69 & 15.06 & 7.14 \\
    \checkmark & \checkmark & & \checkmark & 5(10\%) & 0 & 61.04 & 52.07 & 17.50 & 8.93 \\
    \checkmark & \checkmark & \checkmark & \checkmark & 5(10\%) & 0 & 61.64 & 52.28 & 15.46 & 6.60 \\\hline
    \checkmark & \checkmark & &  & 5(10\%) & 43 & 72.23 & 65.90 &  8.04 & 3.27 \\
    \checkmark & \checkmark & \checkmark &  & 5(10\%) & 43 & 73.58 & 67.33 & 7.45 & 2.90 \\
    \checkmark & \checkmark & & \checkmark & 5(10\%) & 43 & 74.15 & 67.21 & 7.91 & 3.12 \\
    \checkmark & \checkmark & \checkmark & \checkmark & 5(10\%) & 43 & 74.59 & 68.32 & 7.38 & 2.96 \\ \hline
    \end{tabular}
}
\end{table}

\begin{table}[h]
\begin{minipage}[c]{0.47\textwidth}
    \centering
    \caption{Ablation studies for the position point sets involved in ITC.} \label{Ablation Tab ITC}
    \setlength{\tabcolsep}{1.7mm}
    \resizebox{!}{1.25cm}{
    \begin{tabular}{cccccccc}
    \hline
    \multirow{2}[0]{*}{$Z^{K}$}  & \multirow{2}[0]{*}{$Z^{B}$} & \multirow{2}[0]{*}{$\hat{Z}^{K}$} & \multirow{2}[0]{*}{$\hat{Z}^{B}$} & \multicolumn{4}{c}{\textbf{Metric}} \\
    \cmidrule(l){5-8}
    &  &  &  &\multirow{1}{*}{Dice $\uparrow$} & \multirow{1}{*}{Jac $\uparrow$} & \multicolumn{1}{c}{\multirow{1}{*}{HD95 $\downarrow$}} & \multicolumn{1}{c}{\multirow{1}{*}{ASD $\downarrow$}} \\ \hline
    \checkmark & & & & 72.98 & 66.57 & 8.26 & 2.48 \\
    \checkmark & \checkmark & & & 72.88 & 66.90 & 8.12 & 2.66 \\
    \checkmark & \checkmark & \checkmark & & 73.03 & 67.03 & 7.47 & 3.16 \\
    \checkmark & \checkmark & \checkmark & \checkmark & 73.58 & 67.33 & 7.45 & 2.90 \\ \hline
    \end{tabular}
    }
\end{minipage} \hfill
\vspace{3mm}
\begin{minipage}[c]{0.47\textwidth}
    \centering
    \caption{Ablation studies for the position point sets involved in CTC.} \label{Ablation Tab CTC}
    \setlength{\tabcolsep}{2mm}
    \resizebox{!}{1.1cm}{
        \begin{tabular}{ccccccc}
        \hline
        \multirow{2}[0]{*}{$L_{f}$} & \multirow{2}[0]{*}{$\hat{Z}^{K}$} & \multirow{2}[0]{*}{$\hat{Z}^{B}$}  & \multicolumn{4}{c}{\textbf{Metric}} \\
        \cmidrule(l){4-7}
        & & & \multirow{1}{*}{Dice $\uparrow$} & \multirow{1}{*}{Jac $\uparrow$} & \multicolumn{1}{c}{\multirow{1}{*}{HD95 $\downarrow$}} & \multicolumn{1}{c}{\multirow{1}{*}{ASD $\downarrow$}} \\ \hline
        \checkmark & & & 73.56 & 66.65 & 7.77 & 3.52 \\
        \checkmark & \checkmark & & 73.82 & 66.35 & 8.07 & 3.27  \\
        \checkmark & \checkmark & \checkmark & 74.15 & 67.21 & 7.91 & 3.12 \\ \hline
        \end{tabular}
    }
\end{minipage} \hfill
\vspace{-3mm}
\end{table}

\subsubsection{\textbf{Performance comparison}}
The results for the CAA-Seg dataset are presented in Table. \ref{Comparison Tab 1} and for AD-CTA in Table. \ref{Comparison Tab 2}.
The tables demonstrate that our method outperforms these semi-supervised methods and approaches the performance of the fully supervised baseline. We also conduct a comparison with the publicly reported performance of our baseline methods on the ACDC dataset. This serves two main purposes: first, to prove the reliability, and second, to demonstrate the effectiveness of the general medical image segmentation tasks. As the Table. \ref{Comparison Tab ACDC}, Our method has improved performance compared with previous semi-supervised methods. 
\
\subsubsection{\textbf{Ablation studies for the main components}}
We conduct ablation studies, as shown in Table. \ref{Ablation Tab}. The SDT regression task branch improves segmentation performance, and the ITC and CTC further enhance the performance. Finally, the combination of all components yields optimal results, demonstrating their effectiveness in our method. Both the intra- and the cross-frame topological Consistency are good for learning spatial continuity.

\subsubsection{\textbf{Ablation studies for the skeleton/boundary position point sets}}
We also conduct experiments on skeleton/boundary position point sets involved in our two topological constraints, ITC and CTC, shown in Table. \ref{Ablation Tab ITC}. To reiterate, these variables shown as the Table. \ref{Variable Summary}. In summary, these constraint position point sets are effective for both ITC and CTC.

\section{Conclusion}
In this paper, we introduce a novel semi-supervised framework for coronary plaque segmentation that utilizes two key consistency mechanisms: Inter-frame Topological Consistency (ITC) and Cross-frame Topological Consistency (CTC). This dual-consistency approach effectively combines local structural details with spatial continuity, resulting in more accurate and reliable segmentation outcomes. Our experiments show that this method outperforms other semi-supervised techniques in Coronary Atherosclerosis Analysis (CAA) and is also highly effective in broader medical image segmentation tasks.

\section{Acknowledgement}
\vspace{-2mm}
This work was supported by the National Natural Science Foundation of China (No. 62471418), the Fujian Provincial Natural Science Foundation of China (No. 2024J01058), and in partl by Xiamen City Industry-University-Research Subsidy Project (2023CXY0111 and 2024CXY0102).

\bibliographystyle{IEEEtran}
\bibliography{mybibliography}
\end{document}